\documentclass[dvips]{acta}
\usepackage{supertabular,lscape,epsfig}
\usepackage{amssymb}
\usepackage{amsmath}
 \SetPages{0}{0}

\newcommand{\brho}{\boldsymbol{\rho}}

\newcommand{\bomega}{\boldsymbol{\omega}}

\def\expf#1{\ensuremath{\mathrm{e}^{#1}}}

\def\mnras{Monthly Notices of the Royal Astronomical Society}
\def\aap{Astronomy and Astrophysics}

\def\apj{Astrophysical Journal}

\def\prc{Physical Rev. C}

\begin{document}
\begin{Titlepage}
\Title{Observational Tests of Neutron Star Relativistic Mean Field
Equations of State}

\Author{M. U~r~b~a~n~e~c$^1$, E. B~\v{e}~t~\'{a}~k$^{1,2}$, and  Z.
S~t~u~c~h~l~\'{\i}~k$^1$} {$^1$Institute of Physics,  Silesian
           University, 74601 Opava, Czech Republic\\
$^2$ Institute of Physics, Slovak Academy of Sciences, 84511
Bratislava, Slovakia}

%
%
\end{Titlepage}

\Abstract{Set of neutron star observational results is used to test
 some selected equations of state of dense nuclear matter. The first observational result
comes from the mass--baryon number relation for pulsar B of the
double pulsar system J 0737--3039. The second one is based on the
mass--radius relation coming from observation of the thermal
radiation of the neutron star RX J 1856.35--3754. The third one
follows the population analysis of isolated neutron star thermal
radiation sources. The last one is the test of maximum mass. The
equation of state of asymmetric nuclear matter is given by the
parameterized form of the relativistic Brueckner-Hartree-Fock mean
field, and we test selected parameterizations that represent fits of
full relativistic mean field calculation. We show that only one of
them is capable to pass the observational tests. This equation of
state represents the first equation of state that is able to explain
all the mentioned observational tests, especially the very accurate
test given by the double pulsar even if no mass loss is
assumed.}{Stars: neutron, Equations of state, Dense matter}


\section{Introduction}

Neutron stars are compact objects that play important role in
different areas of modern physics. Here we concentrate our attention
on the possibility that phenomena related to neutron stars can be
used as tests of equation of state (EoS) of asymmetric nuclear
matter. The tests used in this paper represent a subset of tests
used previously by Kl\"ahn \etal (2006). We focus our attention on
tests that come from astronomical observations. However we have not
applied the very promising test coming from the observations of
quasiperiodic oscillations (QPOs), since the theory and data
interpretation is still in progress (see e.g. T\"or\"ok \etal
(2008a,b,c, 2010) van der Klis (2004)). The QPO test applied on the
4U 1636--536 object in Kl\"ahn \etal (2006) represents the maximum
mass test in the present paper, and another test following the QPO
phenomena observed in the 4U 0614+09 object do not provide a strong
test and all the EoS tested in this paper pass it.

A wide spectrum of different equations of state of nuclear matter
and their applications to astrophysical problems has been reported
in literature (see, e.g., Haensel, Zdunik and Douchin 2002, Rikovska
Stone \etal 2003, Weber, Negreiros and Rosenfeld 2007, Lattimer and
Prakash 2007, Burgio 2008). Some of the EoS collections (even though
not all of them are up-to-date already) give an amazingly rich
general overview of the state-of-the-art, whereas the others
emphasize some specific aims. All these EoS yield (nearly) the same
properties close to the standard nuclear density ($\rho_N \approx$
0.16 nucleon/fm$^3 \approx 2.7 \times 10^{14}~$g/cm$^3$), but when
one is far off this value, s/he has to rely more on underlying
principles than on possible experimental verification of predicted
physical observables.

Here we concentrate our  attention on relativistic asymmetric
nuclear matter where the EoS stem from an assumed form of the
interaction Lagrangian. The calculations use the relativistic
mean-field theory with allowance for an isospin degree of freedom
(Kubis and Kutchera 1997, M\"uler and Serot 1996). We employed the
Dirac-Brueckner-Hartree-Fock mean-field approach in its
parameterized form suggested in Gmuca (1991) which reproduces the
nuclear matter results of Huber, Weber and Weigel (1995). That has
been used to calculate high-density behavior of asymmetric nuclear
matter with varying neutron-to-proton ratio (Gmuca 1992). The proton
fraction has been determined from the condition of
$\beta$-equilibrium and charge neutrality, and it is
density-dependent. We have extended our calculations for densities
up to $4 \times \rho_N$ and if there was an astrophysical motivation
even higher.

The EoS is used to model the static, spherically symmetric neutron
star in the framework of general relativity. The equation of
hydrostatic equilibrium is solved for different central parameters
(pressure, energy density, baryon number density). The radius of the
neutron star model is then given by the condition of vanishing
pressure. The resulting properties of the neutron star model are
then compared with observational data. From the test ensemble
presented by Kl\"ahn \etal (2006) we choose four astrophysical
observations to test our selected parameterizations that have been
found to be a good description of nuclear matter at subnuclear
densities for pure neutron matter and up to $ 2 \times \rho_N $ for
symmetric nuclear matter (Kotuli\v c Bunta and Gmuca 2003).

The maximum mass test is the standard way to test the EoS of
asymmetric nuclear matter (see e.g. Haensel, Potekhin and Yakovlev
2007, Lattimer and Prakash 2007, Kl\"ahn \etal 2006). The usual
value to constrain the maximal mass of neutron star comes from
observations of double pulsar PSR 0751+1807 giving $M=(2.1 \pm
0.2)~M_\odot$, with $M_\odot$ being the solar mass. This value was,
however, lowered to $M=(1.26 \pm 0.14)~M_\odot$ (Nice, Stairs and
Kasian 2008) and could not be used as maximum mass test anymore.
Another value that could serve as maximum mass test comes from the
observation of QPOs. The mass is constrained on the basis that the
observed frequency corresponds to the frequency at innermost stable
circular orbit (Barret, Olive and Miller 2005, Belloni, Mendez and
Homman 2007, van der Klis 2004).

Popov \etal (2006) used the population synthesis of the isolated
neutron star sources of thermal radiation and concluded that the
neutron stars with mass $M<1.5~M_\odot$ could not cool via direct
URCA reactions. This conclusion follow from the fact that all
observed sources of thermal radiation have masses bellow the quoted
value. This could be explained by the fact that more massive objects
cool via the direct URCA reactions which represents the fast cooling
scenario and thus the thermal radiation could not be detected. These
arguments were used by Kl\"ahn \etal (2006) to build a strong and a
weak test on EoS.

A very accurate test of EoS was developed by Podsiadlowski \etal
(2005). They based it on the model of double pulsar system J
0737--3039 formation. The model predicts the pulsar B of this system
to be born via the electron capture supernova what suggests
extremely low mass loss and thus the number of particles conserved
during the progenitor collapse to neutron star. This put limits on
the mass--baryon number relation. Instead of the baryon number that
represents the total number of baryons contained in the neutron
star, the baryon mass could be used equally.

The thermal radiation coming from the neutron star source RX
J1856.35--3754 could be used to put limits on the mass--radius
relation of the neutron star model. Tr\"umper \etal (2004) used two
different models to explain the spectral feature for this specific
source and found its apparent radius that represents the radius of
the neutron star as seen by a distant observer. The analyses of data
to obtain the isolated neutron star radius strongly depend on the
radiation spectrum emitted by the object and the estimated radius is
proportional to the distance from Earth to the source. The distances
obtained for RX J1856.5-3754 range from $D=61^{+9}_{-8}$~pc (Walter
and Matthews 1997) to $D=161^{+18}_{-14}~$pc (van Kerkwijk and
Kaplan 2007). The derived apparent radius $R_\infty$ is given by the
model of the atmosphere. The original model by Pons \etal (2002)
resulted in $R_\infty / D = 0.13$~km.pc$^{-1}$. Tr\"umper \etal
(2004) presented new models of atmosphere leading to the estimates
of $R_\infty=16.5$~km for the two component model of spectra and
$R_\infty=16.8$~km assuming continuous temperature distribution
model. If the distance derived by van Kerkwijk and Kaplan (2007) and
the original model of Pons \etal (2002) are used together, they lead
to unexpectedly high estimate $R_\infty = 20.9$~km. Recently
Steiner, Lattimer and Brown (2010) presented results based on new
analysis of data giving the distance $119 \pm 5$~pc and the original
model for atmosphere (Pons \etal 2002) then implies
$R_\infty=15.47$~km. We decided to use the three values
$R_\infty=15.5,~16.8,~20.9$~km to put limits on neutron star
equation of state.

Another promising way to constrain the equation of state are the
moment of inertia measurements (see e.g. Lattimer and Prakash 2007
and references therein). Two ways have been proposed quite recently.
One for the Crab pulsar (Bejger and Haensel 2002,2003) following
observations of the pulsar-nebula system, and the other for the
pulsar A of the double pulsar system J0737--3039 (Bejger, Bulik and
Haensel 2005) based on the measurements of the second order post
Newtonian parameters of the binary system. Even thought both ways
could provide strong limits on the equations of state in principle,
they need more accurate observational inputs. We need better
estimates of the mass of Crab nebula in the first case and very
accurate measurements of orbital parameters are necessary to
calculate the moment of inertia in the second case. For these
reasons we do not include these tests to our calculations. The
measurements of moment of inertia of the neutron star together with
its mass put limits on the radius of the neutron star that is
crucial for the cooling scenarios (see e.g. Lattimer and Prakash
2007, Stuchl\'\i k \etal 2009).

The paper is organized as follows. In section\,2 we present our EoS
and details of the neutron star matter description. Section\,3
briefly summarizes the model of static spherically symmetric neutron
star. We present our results and compare them to observations in
section\,4. The paper is closed by conclusions in section\,5.

%
\section{Equation of state of neutron star matter}

\subsection{Asymmetric nuclear matter in relativistic mean-field
approach}

We follow the Dirac-Brueckner-Hartree-Fock (DBHF) mean field (see
Weber 1999, Walecka 2004, de Jong and Lenske 1998, Krastev and
Sammarruca 2006 for underlying theories), which easily allows to
consider different neutron-proton composition of the neutron star
matter, and also the inclusion of non-nucleonic degrees of freedom.

The full mean-field DBHF calculations of nuclear matter (Huber,
Weber and Weigel 1995, Lee \etal 1998, Li, Machleidt, and Brockmann
1992) have been parameterized by Kotuli\v c Bunta and Gmuce (2003),
and we employ their parameterization with one-boson-exchange (OBE)
potential A of Brockmann and Machleidt Li, Machleidt, and Brockmann
 (1992). We refer to the paper of Kotuli\v c Bunta and Gmuca (2003) for the
explicite set of values of the corresponding parameters. The model
Lagrangian density includes the nucleon field $\psi$, isoscalar
scalar meson field $\sigma$, isoscalar vector meson field $\bomega$,
isovector vector meson field $\brho$ ,and isovector scalar meson
field $\delta$, including also the vector cross-interaction. The
Lagrangian density in the form used by Kotuli\v c Bunta and Gmuca
(2003) reads
\begin{eqnarray}
  {\cal L} (\psi, \sigma, \bomega, \brho, \delta) =
  \bar{\psi} [ \gamma_\mu (i \partial^\mu - g_\omega \bomega^\mu)
                           - (m_\mathrm N -g_\sigma \sigma)] \psi
  \nonumber \\
  + \frac{1}{2} (\partial_\mu \sigma \partial^\mu \sigma
      - {m_\sigma}^2 \sigma^2)
      - \frac{1}{4}\bomega_{\mu \nu} \bomega^{\mu \nu}
      + \frac{1}{2} {m_\omega}^2 \bomega_\mu \bomega^\mu
  \nonumber \\
  - \frac{1}{3} b_\sigma m_\mathrm N (g_\sigma \sigma)^3
      - \frac{1}{4} c_\sigma (g_\sigma \sigma)^4
      + \frac{1}{4} c_\omega ({g_\omega}^2 \bomega_\mu \bomega^\mu)^2
  \nonumber \\
  + \frac{1}{2} (\partial_\mu \delta \partial^\mu \delta
                          - {m_\delta}^2 \delta^2)
      + \frac{1}{2} {m_\rho}^2 \brho_\mu \brho^\mu
      - \frac{1}{4} \brho_{\mu \nu} \brho^{\mu \nu}
  \nonumber \\
  + \frac{1}{2} \Lambda_V ({g_\rho}^2 \brho_\mu \brho^\mu)
                              ({g_\omega}^2 \bomega_\mu
  \bomega^\mu)
      - g_\rho \brho_\mu \bar{\psi} \gamma^\mu \tau \psi
      + g_\delta \delta \bar{\psi} \tau \psi ,
\end{eqnarray}
where the antisymmetric tensors are
\begin{eqnarray}
   \bomega_{\mu \nu}  & \equiv & \partial_\nu \bomega_\mu
                         - \partial_\mu \bomega_\nu ,\nonumber \\
   \brho_{\mu \nu} & \equiv & \partial_\nu \brho_\mu
                         -  \partial_\mu \brho_\nu ;
\end{eqnarray}
the strength of the interactions of isoscalar and isovector mesons
with nucleons is given by (dimensionless) coupling constants $g$'s
and the self-coupling constants (also dimensionless) are $b_\sigma$
(cubic), $c_\sigma$ (quartic scalar) and $c_\omega$ (quartic
vector). The second and the fourth lines represent non-interacting
Hamiltonian for all me\-sons, $\Lambda_V$ is the cross-coupling
constant of the interaction between $\bomega$ and $\brho$ mesons.
Furthermore, $m_\mathrm N$ is the nucleon mass, $\partial^\mu \equiv
\frac{\partial}{\partial x_\mu}$ and $\gamma$'s are the Dirac
matrices (Kotuli\v c Bunta and Gmuca 2003, Serot and Walecka 1986,
Weber 1999).


We choose here three following parameterizations, which were shown
to yield the best fits to the well-known properties of nuclear
matter
\begin{itemize}
\item[$H$] HA in Kotuli\v c Bunta and Gmuca (2003) represents the best RMF fit to
results obtained by Huber, Weber and Weigel 1995.
\item[$L$] LA in Kotuli\v c Bunta and Gmuca (2003) represents the best RMF fit to
results obtained by Lee \etal 1998, but does not include the
$\delta$ mesons to nucleons coupling.
\item[$M$] MA in Kotuli\v c Bunta and Gmuca (2003) represents the best RMF fit to
results obtained by Li, Machleidt, and Brockmann 1992, but does not
include the $\delta$ mesons to nucleons coupling.
\end{itemize}

 The EoS of Kotuli\v c Bunta and Gmuca which have been found to be a good description
of asymmetric nuclear matter, are easily expressed up to about $ 4
\times \rho_N $ (parameterization $H$) or even higher
(parameterizations $L$ and $M$).

\subsection{$\beta$-equilibrium}

The total energy density of {\it{n-p-e-$\mu$}} matter is given by
\begin{equation}
\mathcal E = \mathcal E_\mathrm B (n_\mathrm B, x_\mathrm p) +
               \mathcal E_e(n_e) +  \mathcal E_\mu(n_\mu),
\end{equation}
where  $\mathcal E_\mathrm B (n_\mathrm B, x_\mathrm p)$ is the
binding energy density of asymmetric nuclear matter, $n_i$ is the
number density of different particles ($i=$ n, p, e, $\mu$),
$n_\mathrm B= n_\mathrm p +n_\mathrm n$ is the baryon number density
and $x_\mathrm p = n_\mathrm p/ n_\mathrm B$ is the proton fraction.
The leptonic contributions $\mathcal E_l(n_l)$ ($l=$ e, $\mu$) to
the total energy density are given by
\begin{equation}
\mathcal E_l (n_l)= \frac{2}{h^3} \int\limits_0^{p_{F(l)}} \left(
m_l^2 c^4 + p^2 c^2 \right )^{1/2}4 \pi p^2 \mathrm d p,
\end{equation}
where $p_{F(l)}$ is the Fermi momentum of {\it l-}th kind of
particle.

The matter in neutron stars is in $\beta$-equilibrium, i.e. in
equilibrium with respect to $ \mathrm{n} \leftrightarrow \mathrm{p}
+ \mathrm{e}^- \leftrightarrow \mathrm{p} + \mu$. The
(anti)neutrinos contribution could be neglected, because the matter
is assumed to be cold enough that they can freely escape. The
equilibrium is given by equality of chemical potentials $
\mu_\mathrm n=\mu_\mathrm p + \mu_\mathrm e = \mu_\mathrm p +
\mu_\mu$, where the chemical potential of each kind of particle is
given by $ \mu_i= {\partial \mathcal E} / {\partial n_i}$. The
chemical potentials of electrons and muons are simply $ \mu_\mathrm
l = \sqrt{m_\mathrm l^2 c^4 + p_{F(\mathrm e)}^2 c^2} $,
while the chemical potentials of nucleons are
\begin{equation}
\mu_\mathrm{(p,n)} =\frac{\partial }{\partial n_\mathrm{(p,~n)}
}\left(\mathcal E_\mathrm B \right) .
\end{equation}

The binding energy density of asymmetric nuclear matter could be
expressed in terms of proton fraction $x_\mathrm p$ (Danielewicz and
Lee 2009)
\begin{equation}
\mathcal E_\mathrm B (n_\mathrm B, x_\mathrm p) = \mathcal
E_\mathrm{SNM}(n_\mathrm B) + (1-2 x_\mathrm p)^2 S(n_\mathrm B),
\label{bindEn}
\end{equation}
where $\mathcal E_\mathrm{SNM}$ is the energy density of symmetric
nuclear matter ($x_\mathrm p=0.5$) and $S(n_\mathrm B)$ is the
symmetry energy density, that corresponds  to the difference of
binding energy density between pure nuclear matter and symmetric
nuclear matter

The symmetry energy $S(n_\mathrm B)$ is the factor corresponding to
the second order term in expansion of binding energy density in
terms of asymmetry parameter $\delta = (n_\mathrm n - n_\mathrm
p)/(n_\mathrm n + n_\mathrm p)= 1 - 2 x_\mathrm p$ and reads
\begin{equation}
S(n_\mathrm B)= \frac{1}{2} \left. \frac{\partial ^2 \mathcal
E_\mathrm B (n_\mathrm B, \delta)}{\partial
\delta^2}\right|_{\delta=0}.
\end{equation}
From equation (6)  one can see that symmetry energy is the
difference of binding energy per particle between pure nuclear
matter and symmetric nuclear matter.
\begin{equation}
S(n_\mathrm B)=\mathcal E_\mathrm B (n_\mathrm B, x_\mathrm p=0) -
\mathcal E_\mathrm B (n_\mathrm B, x_\mathrm p=0.5).
\end{equation}
The condition of $\beta$-equilibrium then reads
\begin{eqnarray}
\mu_\mathrm e= \mu_\mu = \mu_\mathrm n - \mu_\mathrm p =
   4\frac{S(n_\mathrm B)}{n_\mathrm B}\left(1-2x_\mathrm p\right).
\end{eqnarray}
and it is solved together with condition of charge neutrality
($n_\mathrm p= n_\mathrm e + n_\mu$) to obtain the proton fraction
of neutron star matter. The binding energy per baryon in dependence
on the baryon number density is illustrated in Figure\,1. The proton
fraction of matter at the beta-equilibrium is given, for the chosen
three EOS parameterizations, as a function of the baryon number
density depicted in Figure\,2.

\subsection{EoS for low densities}

The nuclear EoS have been the dominant input for the calculations in
the high-density region, namely $\rho \geq 10^{14}~\mathrm{g/cm^3}$.
For lower densities, the EoS used are the following:
\begin{itemize}
\item
Feynman-Metropolis-Teller EoS for
  $7.9 ~\mathrm{g/cm^3} \leq \rho \leq 10^4~\mathrm{g/cm^3}$
  where matter consists of
  $\mathrm{e}^-$
  and $^{56}_{26}\mathrm{Fe}$,  Feynman, Metropolis and Teller (1949);
\item
Baym-Pethick-Sutherland EoS for
  $10^4 ~\mathrm{g/cm^3} \leq \rho \leq 4.3 \times 10^{11}~\mathrm{g/cm^3}$
  with Coulomb lattice energy corrections Baym, Pethick, and Sutherland (1971);
\item
Baym-Bethe-Pethick EoS for
  $4.3 ~\mathrm{g/cm^3} \times 10^{11} \leq \rho \leq
  10^{14}~\mathrm{g/cm^3}$: here, e$^-$, neutrons
  and equilibrated nuclei calculated using the compressible liquid
  drop model Baym, Bethe, and Pethick (1971).
\end{itemize}

\begin{figure}[t]
\begin{center}
\includegraphics[width=\textwidth]{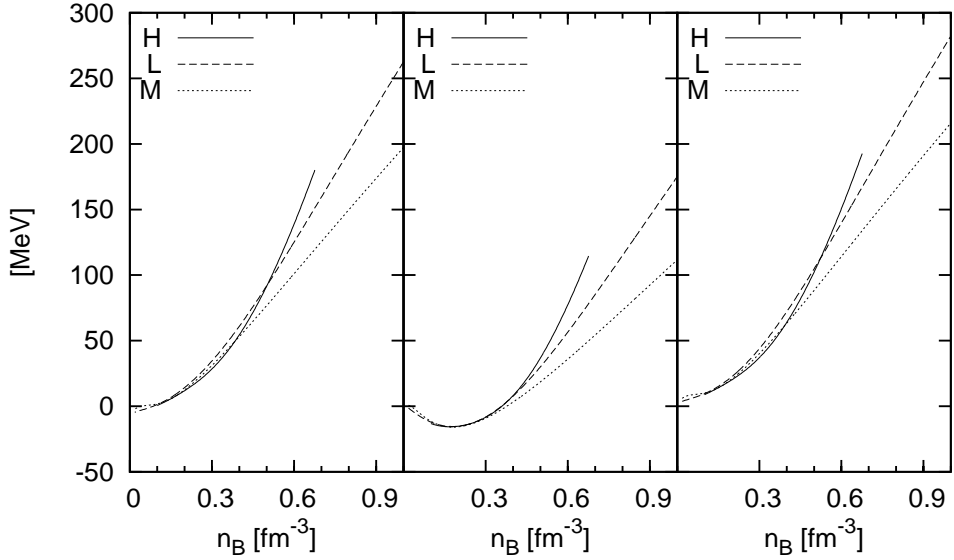}
\end{center}
\caption{\label{E-nb} Binding energy per particle of different types
of nuclear matter for used parameterizations. {\it Left} Matter at
$\beta$--equilibrium, {\it Middle} symmetric nuclear matter, and
{\it Right} pure neutron matter}
\end{figure}

\section{Neutron star models}
We consider static spherically symmetric models of neutron stars.
The interior spacetime is described by the internal Schwarzschild
metric (see, e.g., Misner, Thorne and Wheeler 1973, Haensel,
Potekhin and Yakovlev 2007) that can be written in geometrical units
($c=G=1$) as
\begin{equation}
\mathrm d s^2=-\expf{2\nu} \mathrm d t^2 + \expf{2\lambda}\mathrm d
r^2 + r^2(\mathrm d \theta^2 + \sin^2 \theta \mathrm d \phi^2),
\end{equation}
where the radial component of metric can be expressed as a function
of energy density $\rho$
\begin{equation}
\expf{2\lambda}=\frac{r}{r-2m(r)},~~~~m(r)=4\pi\int_0^r \rho r_1^2
\mathrm d r_1.
\end{equation}

The matter is assumed to be perfect fluid described by the energy
momentum tensor
\begin{equation}
T^{\mu\nu}=(P + \rho)u^\mu u^\nu + P g^{\mu\nu},
\end{equation}
(Misner, Thorne, and Wheeler 1973). where $P$ is the pressure,
$u^\mu$ is the $4$-velocity of matter and $g^{\mu\nu}$ is the metric
tensor. The energy momentum tensor satisfies the conservation law
$T^{\mu\nu}_{~~~;\nu}=0$.

\begin{figure}[t]
\begin{center}
\includegraphics[width=\textwidth]{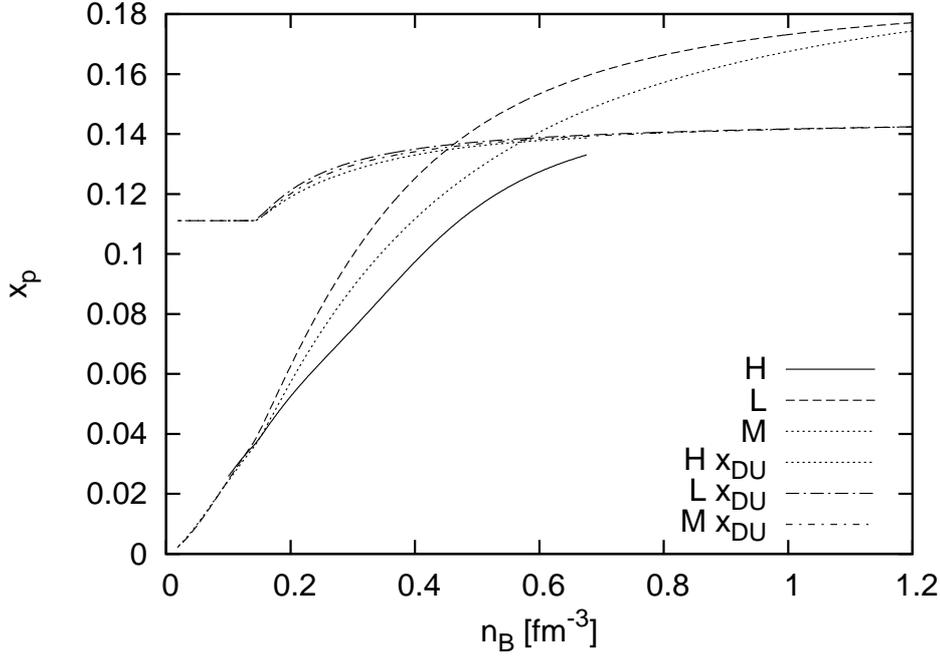} \end{center}
\caption{\label{xp-nb} Proton fraction of matter being at $\beta$ -
equilibrium for used parameterizations. Also lines of direct URCA
threshold (marked with $x_\mathrm{DU}$) for all parameterizations
are depicted.}
\end{figure}

The hydrostatic equilibrium is in general relativity given by the
Tolman-Oppen\-heimer-Volkoff equation (TOV) (Oppenheimer and Volkoff
1939, Tolman 1939), which reads
\begin{equation}
\frac{{\rm d}P}{{\rm d}r}=-(\rho + P)\frac{m(r)+4\pi r^3
P}{r(r-2m(r))}.
\end{equation}
Integration of TOV starting from given central energy density
$\rho_{\rm c}$ uses the EoS and finally yields the radius $R$, given
by the boundary condition $P(R)=0$, and the gravitational mass $M =
m(R)$ of the neutron star.

Another useful quantity to calculate is the so-called baryonic mass
$M_\mathrm B$ that represents the total number of baryons contained
in the neutron star multiplied by the atomic mass unit u. The
baryonic mass is then expressed as
\begin{equation}
M_\mathrm B= 4 \pi \mathrm u \int \limits_0^R n_\mathrm B (r) \left
[ 1 - \frac{2 m(r)}{r} \right]^{-1/2} r^2 \mathrm d r,
\end{equation}
where $n_\mathrm B (r)$ is the baryon number density at the radius
$r$.



%
%
\begin{figure}[t]
\begin{center}
\includegraphics[width=\textwidth]{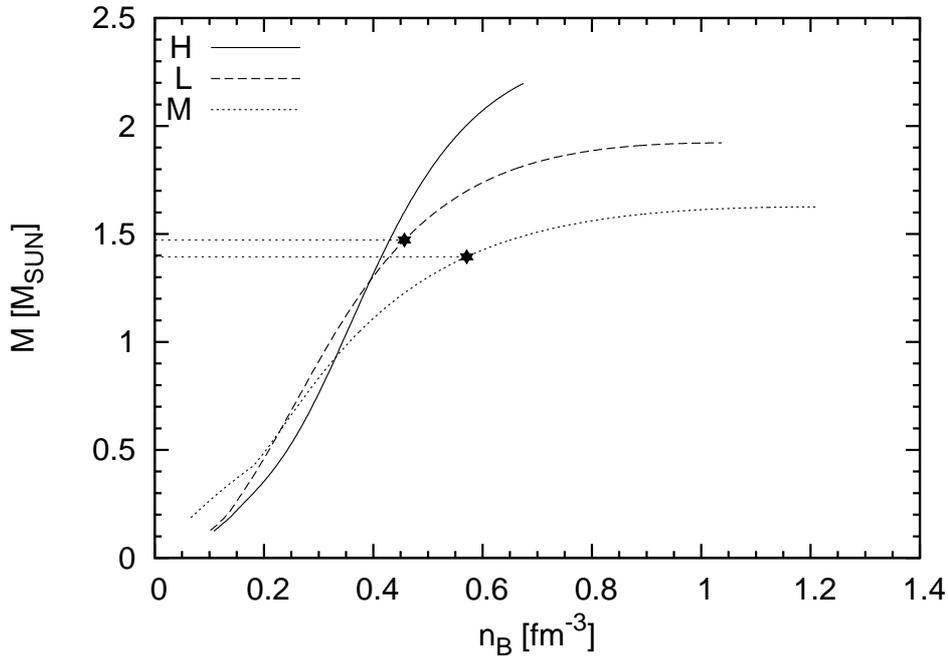}
\end{center}
\caption{\label{M-nb}Mass given as a function of central baryon
number density for different parameterizations. The stars correspond
to the minimum mass of a neutron star that could cool via direct
URCA reactions.}
\end{figure}

\section{Results versus observations}
Several dozens of neutron stars and/or similar objects have their
masses reported; a great majority of them is in very close vicinity
of 1.4~$M_\odot$, and only very few are significantly above (see,
e.g., the compilations in Bethe, Brown and Lee (2007), Lattimer and
Prakash 2007) and observations and analyses (see, e.g, Rikovska
Stone \etal 2003, Weber, Negreiros and Rosenfeld 2007, Podsiadlowski
\etal 2005, Tr\"umper \etal 2004, Pons \etal 2002, Kramer and Wex
2009, Krastev and Sammarruca 2006, Lattimer and Prakash 2007,
Blaschke, Kl\"ahn and Sandin 2008, Dexheimer, Vasconcellos and
Bodmann 2008, Kl\"ahn \etal 2006, Nice, Stairs and Kasian 2008,
Rikovska Stone \etal 2007).
 However, recent results of the
data fitting of kHz quasiperiodic oscillations observed in the
low-mass X-ray systems containing neutron stars indicate relatively
high masses of $M>2~M_\odot$ (Belloni, Mendez and Homan 2007,
T\"or\"ok \etal 2008a,b,c, Barret, Olive and Miller 2005, Boutelier
\etal 2010, Boutloukos \etal 2006) which could provide very strong
constraint on the EoS. On the other hand, modification of the
characteristic orbital frequencies by a magnetic repulsion caused by
the interaction of slightly charged matter in accretion disc in
vicinity of a neutron star with dipole magnetic field could shift
the mass estimates to lower values close to canonical 1.4~$M_\odot$
(Bakala \etal 2008). Our calculations with parameterization $H$
allow for the existence of neutron stars even for so heavy masses.

\subsection{Direct URCA constraints}
The proton fraction $x_\mathrm p$ of matter in $\beta$ equilibrium
is presented in Figure\,2 together with the direct URCA threshold.
The direct URCA reactions $n \rightarrow p + e^- + \bar{\nu}_e$
could operate only if the proton fraction exceeds the threshold
given by the condition
\begin{equation}
x_\mathrm{DU}=\frac{1}{1+\left(1 + x^{1/3}_\mathrm e\right)^3},
\end{equation}
where $x_\mathrm e=n_\mathrm e/(n_\mathrm e + n_\mu)$. One can see
that only parameterizations $L$ and $M$ enable rapid cooling. The
threshold densities are $n_\mathrm{DU} = 0.457~\mathrm{fm}^{-3}$ in
the case of parameterization $L$ and  $n_\mathrm{DU} =
0.571~\mathrm{fm}^{-3}$ in the case of parameterization $M$. These
values correspond (see Figure\,3) to neutron star masses
$M=1.47~M_\odot$ (parameterization $L$) and $M=1.39~M_\odot$
(parameterization $M$). Parameterizations $L$ and $M$ thus do not
fulfill the direct URCA constraints, however Kl\"ahn \etal 2006 used
also the value $1.35~M_\odot$ as a weaker test that is passed also
by parameterizations $L$ and $M$.

\subsection{Maximum mass}
The maximum mass limit is probably the most often used test of the
equation of state. The maximum masses given by EoS used in this
paper are $M^H_\mathrm{max} = 2.18~M_\odot$, $M^L_\mathrm{max} =
1.92~M_\odot$ and $M^M_\mathrm{max} = 1.62~M_\odot$. The maximum
mass obtained for objects containing matter described by
parameterizations $L$ and $M$ follows the requirements of stability
with respect to radial oscillations ($\partial M/\partial n_c >0$).
In the case $H$ we used the values corresponding to the central
density $n_\mathrm B (r=0) = 0.66~\mathrm{fm}^{-3}$, because for the
densities above this value the model used for the EoS is not without
questions and also because only with central densities up to about
$4\times$ normal nuclear density we were able to explain masses of
neutron stars that meet the observational requirements. With some
extrapolations, higher masses could be in principle modelled, but we
decided to use parameterization $H$ up to the quoted density only
since there is no current astrophysical observation of such a high
mass. The observation of high mass is however crucial and very
promising issue of astrophysical observations. It should be noted
that Kl\"ahn \etal 2006 used the value that could not be used
anymore. Also the result for the source 4U 1636--536 that gives $M =
(1.9 - 2.1)~M_\odot $ as proposed by Barret, Olive and Miller (2005)
should be used rather as an upper limit of the neutron star mass
than as its estimate, see, e.g. Miller, Lamb and Psaltis (1998) for
underlying theories. The neutron star mass is inferred due to the
highest observed frequency of QPOs observed in the system, under the
assumption of identifying the highest frequency with the Keplerian
frequency of the innermost stable circular orbit (ISCO). Clearly
this gives an upper limit on the mass, and the real neutron star
mass has to be expected smaller because the QPOs have to be excited
above ISCO. Up to date, one of the two pulsars Ter 5 I and J has a
reported mass larger than $1.68~M_\odot$ to 95\% confidence level
(see, e.g. Lattimer and Prakash (2007) and references
therein)\footnote{The individual pulsar masses unfortunately are not
assumption-independent. In our discussion, we adhere to the value
$1.68~M_\odot$ reported by Lattimer and Prakash, but bearing in mind
the possible uncertainty in its derivation.}.  Champion \etal (2008)
predicted mass of PSR J1903+0327 to be $M=1.74~\pm 0.04 M_\odot$.
Freire (2009) estimated the mass for the same source to be $M = 1.67
\pm 0.01~M_\odot$. These values, even if they are different, give
approximately the same limit on mass when they are combined
together, namely $\gtrsim 1.66~M_\odot$ at 2 $\sigma$ level. These
predictions are not in favour the parameterization $M$ with
$M^M_\mathrm{max} = 1.62~M_\odot$.

\begin{figure}[t]
\begin{center}
\includegraphics[width=\textwidth]{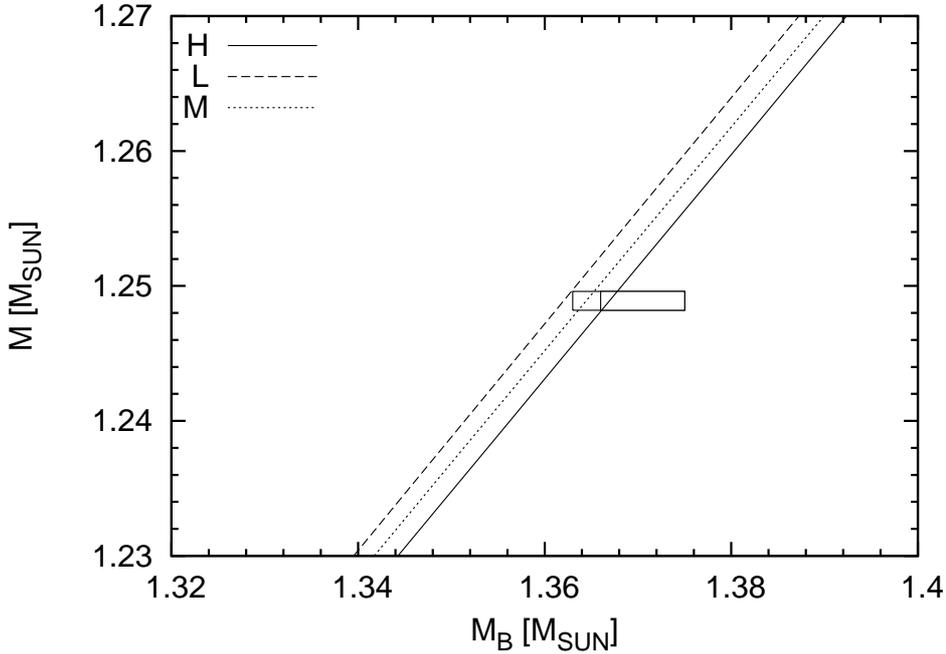}
\end{center}
\caption{\label{M-Mb}Relation of calculated gravitational mass $M$
and the baryonic one $M_\mathrm B$ for different parameterizations.
The limitations imposed by the analysis of the J0737-3039 double
pulsar are drawn as a small rectangle. The box is also extended to
the left by $0.003 M_\odot$ indicating the possible mass loss.}
\end{figure}

\subsection{Double pulsar J0737--3039}

Podsiadlowski \etal (2005) investigated possible formation scenarios
of double pulsar J0737--3039. They have shown that one can test EoS
assuming the pulsar B is formed by an electron-capture supernova.
Such scenario enables formation of the pulsar B that has low but
very accurately measured mass $M=1.2489 \pm 0.0007~ M_\odot$ (Kramer
and Wex 2009). If this pulsar is born under the presented scenario,
its baryonic mass $M_\mathrm B$ should be in the range 1.366 to
1.375~$M_\odot$. The authors also argue the matter loss being low
(the matter loss they give is few times $10^{-3}~M_\odot$). The
relation between the gravitational and the baryonic masses together
with the limitations derived from the double pulsar observations are
presented in Figure\,4. One can see that the only parameterization
that meets requirements assuming no mass loss is the
parameterization $H$. The parameterization $M$ is able to explain
the results if one includes mass loss predicted by Podsiadlowski
\etal (2005). Unfortunately this parameterization was ruled out by
the maximum mass test.

\begin{figure}[t]
\begin{center}
\includegraphics[width=\textwidth]{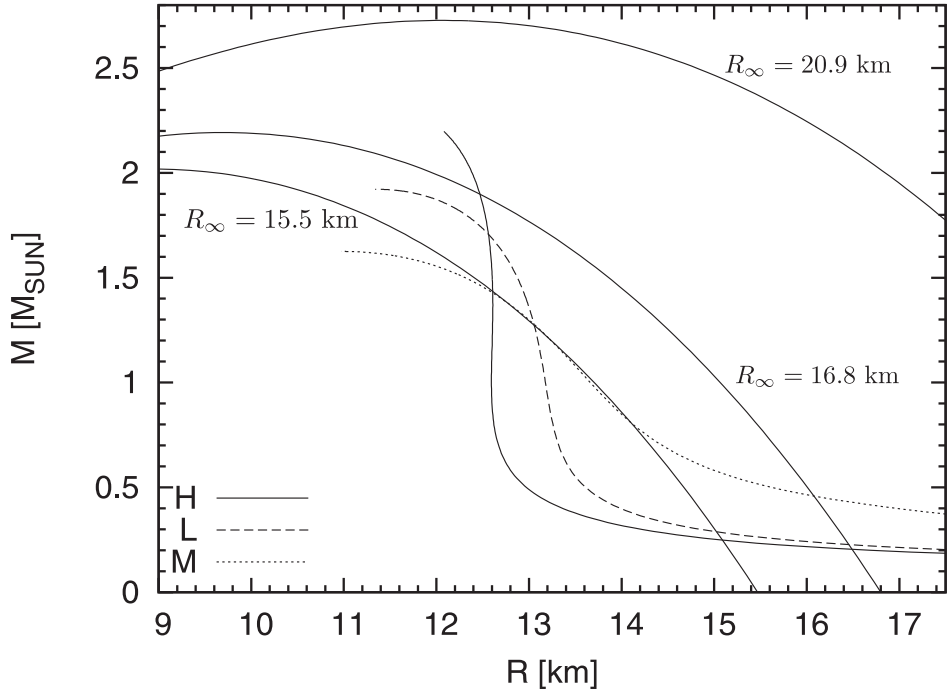}
\end{center}
\caption{\label{M-R}Mass--radius relation for different
parameterizations. The lines corresponding to RX J 1856.5--3754
gives a lower mass limit, that should the given EoS get over.}
\end{figure}

\subsection{Isolated neutron star RX J1856.5--3754}

Several authors (see, e.g., Tr\"umper \etal 2004, Pons \etal 2002,
van Kerkwijk and Kaplan 2007, Steiner \etal 2010) discussed
observations of the isolated neutron star RX J1856.5--3754 and they
found constraints on the mass-radius relation of this particular
neutron star. They found the limits of the apparent radius being
given byt the mass-radius relation
\begin{equation}
\frac{M}{M_\odot}=\frac{R}{2.95~{\mathrm{km}
}}\left(1-\frac{R^2}{R^2_\infty}\right),
\end{equation}
that could serve as a test of equation of state. We have used three
different values for $R_\infty$ namely $R_\infty =
15.5,~16.8,~20.9$~km. None of tested parameterizations is able to
explain the apparent radius $R_\infty = 20.9$~km. Parameterization
$H$ is the only one capable of explaining the apparent radius
$R_\infty = 16.8$~km estimated by Tr\"umper \etal (2004). The lowest
predicted apparent radius could be modeled by all parameterizations
considered in this paper. The mass--radius relations for all
parameterizations together with observational limits are illustrated
in Figure\,5.

\section{Conclusions}
We have employed the parameterized form of the relativistic
mean-field EoS for asymmetric nuclear matter with vector cross
interaction. The proton fraction was varied in accord with the need
of the $\beta$-equilibrium and charge neutrality. Assuming
spherically symmetric geometry and using TOV equation, we
constructed models of neutron stars for different central
parameters. We have used set of observational data to test EoS of
nuclear matter represented by three different parameterizations of
relativistic Brueckner-Hartree-Fock equation.

We have shown that only the parameterization $H$ is able to pass
almost all the tests considered in this paper. The only exception is
the apparent radius $R_\infty = 20.9$~km estimation for the isolated
neutron star RX J1856.5-3754; however this estimate is based on
distance measurements being still widely discussed. This
parameterization also represents the only EoS based on the
relativistic Brueckner-Hartree-Fock theory that could explain the
formation of pulsar B in the double pulsar system J 0737--3039
without mass loss.

Our present calculations have been done considering only neutrons
and protons in $\beta$-equilibrium with electrons and muons. We aim
to continue in tests of given EoS in future. One of our plans is to
include hyperons. Another is to perform more detailed tests based on
the promising fitting of observational data of quasi-periodic
oscillations in low-mass X-ray systems measurements.  This
necessitates to investigate the rotational effects on neutron star
models based on the Hartle-Thorne metric reflecting mass, spin and
the quadrupole moment of the neutron star (Hartle 1967, Hartle and
Thorne 1968). Our preliminary results indicate that these
improvements could bring a new information on the validity of EoS
(Stuchl\'\i k \etal 2007). The important role of the neutron star
spin is demonstrated in the case of Circinus X--1 (T\"or\"ok \etal
2010).


\Acknow{The authors are grateful to J. Kotuli\v c Bunta and \v S.
Gmuca for the availability of their computer codes and to F. Weber
for sending some of his collected data. The work has been supported
by the Czech grants MSM 4781305903 (EB and ZS) and LC 06014 (MU) and
by the VEGA grant 2/0029/10 (EB). One of the authors (ZS) would like
express his gratitude to Czech Committee for Collaboration with CERN
for support and the CERN Theory Division for perfect hospitality.
Authors are also grateful to anonymous referee for his/her comments
and suggestions.}

\end{document}